\documentclass[10pt,a4paper,onecolumn]{article}
\usepackage{marginnote}
\usepackage{graphicx}
\usepackage{xcolor}
\usepackage{authblk,etoolbox}
\usepackage{titlesec}
\usepackage{calc}
\usepackage{tikz}
\usepackage{hyperref}
\hypersetup{colorlinks,breaklinks,
            urlcolor=[rgb]{0.0, 0.5, 1.0},
            linkcolor=[rgb]{0.0, 0.5, 1.0}}
\usepackage{caption}
\usepackage{tcolorbox}
\usepackage{amssymb,amsmath}
\usepackage{ifxetex,ifluatex}
\usepackage{seqsplit}
\usepackage{fixltx2e} 
\usepackage[
  backend=biber,
]{biblatex}
\bibliography{references.bib}

\usepackage[top=3.5cm, bottom=3cm, right=1.5cm, left=1.0cm,
            headheight=2.2cm, reversemp, includemp, marginparwidth=4.5cm]{geometry}

\titleformat{\section}
  {\normalfont\sffamily\Large\bfseries}
  {}{0pt}{}
\titleformat{\subsection}
  {\normalfont\sffamily\large\bfseries}
  {}{0pt}{}
\titleformat{\subsubsection}
  {\normalfont\sffamily\bfseries}
  {}{0pt}{}
\titleformat*{\paragraph}
  {\sffamily\normalsize}

\usepackage{fancyhdr}
\makeatletter
\let\ps@plain\ps@fancy
\fancyheadoffset[L]{4.5cm}
\fancyfootoffset[L]{4.5cm}

\definecolor{linky}{rgb}{0.0, 0.5, 1.0}

\newtcolorbox{repobox}
   {colback=red, colframe=red!75!black,
     boxrule=0.5pt, arc=2pt, left=6pt, right=6pt, top=3pt, bottom=3pt}

\newcommand{\ExternalLink}{%
   \tikz[x=1.2ex, y=1.2ex, baseline=-0.05ex]{%
       \begin{scope}[x=1ex, y=1ex]
           \clip (-0.1,-0.1)
               --++ (-0, 1.2)
               --++ (0.6, 0)
               --++ (0, -0.6)
               --++ (0.6, 0)
               --++ (0, -1);
           \path[draw,
               line width = 0.5,
               rounded corners=0.5]
               (0,0) rectangle (1,1);
       \end{scope}
       \path[draw, line width = 0.5] (0.5, 0.5)
           -- (1, 1);
       \path[draw, line width = 0.5] (0.6, 1)
           -- (1, 1) -- (1, 0.6);
       }
   }

\patchcmd{\@maketitle}{center}{flushleft}{}{}
\patchcmd{\@maketitle}{center}{flushleft}{}{}
\patchcmd{\@maketitle}{\LARGE}{\LARGE\sffamily}{}{}
\def\maketitle{{%
  \renewenvironment{tabular}[2][]
    {\begin{flushleft}}
    {\end{flushleft}}
  \AB@maketitle}}
\makeatletter
\renewcommand\AB@affilsepx{ \protect\Affilfont}
\renewcommand\AB@affilnote[1]{{\bfseries #1}\hspace{3pt}}
\makeatother

\renewcommand\Affilfont{\sffamily\small\mdseries}
\ifnum 0\ifxetex 1\fi\ifluatex 1\fi=0 
  \usepackage[T1]{fontenc}
  \usepackage[utf8]{inputenc}

\else 
  \ifxetex
    \usepackage{mathspec}
  \else
    \usepackage{fontspec}
  \fi
  \defaultfontfeatures{Ligatures=TeX,Scale=MatchLowercase}

\fi
\IfFileExists{upquote.sty}{\usepackage{upquote}}{}
\IfFileExists{microtype.sty}{%
\usepackage{microtype}
\UseMicrotypeSet[protrusion]{basicmath} 
}{}

\usepackage{hyperref}
\hypersetup{unicode=true,
            pdftitle={LocalCop: An R package for local likelihood inference for conditional copulas},
            pdfborder={0 0 0},
            breaklinks=true}
\usepackage{graphicx,grffile}
\makeatletter
\def\maxwidth{\ifdim\Gin@nat@width>\linewidth\linewidth\else\Gin@nat@width\fi}
\def\maxheight{\ifdim\Gin@nat@height>\textheight\textheight\else\Gin@nat@height\fi}
\makeatother
\setkeys{Gin}{width=\maxwidth,height=\maxheight,keepaspectratio}
\IfFileExists{parskip.sty}{%
\usepackage{parskip}
}{
\setlength{\parindent}{0pt}
\setlength{\parskip}{6pt plus 2pt minus 1pt}
}
\ifx\paragraph\undefined\else
\let\oldparagraph\paragraph
\renewcommand{\paragraph}[1]{\oldparagraph{#1}\mbox{}}
\fi
\ifx\subparagraph\undefined\else
\let\oldsubparagraph\subparagraph
\renewcommand{\subparagraph}[1]{\oldsubparagraph{#1}\mbox{}}
\fi

\usepackage{color}
\usepackage{fancyvrb}

\DefineVerbatimEnvironment{Highlighting}{Verbatim}{commandchars=\\\{\}}
\usepackage{framed}
\definecolor{shadecolor}{RGB}{248,248,248}
\newenvironment{Shaded}{\begin{snugshade}}{\end{snugshade}}

\newcommand{\AttributeTok}[1]{\textcolor[rgb]{0.13,0.29,0.53}{#1}}

\newcommand{\CommentTok}[1]{\textcolor[rgb]{0.56,0.35,0.01}{\textit{#1}}}

\newcommand{\ControlFlowTok}[1]{\textcolor[rgb]{0.13,0.29,0.53}{\textbf{#1}}}

\newcommand{\DecValTok}[1]{\textcolor[rgb]{0.00,0.00,0.81}{#1}}

\newcommand{\FloatTok}[1]{\textcolor[rgb]{0.00,0.00,0.81}{#1}}
\newcommand{\FunctionTok}[1]{\textcolor[rgb]{0.13,0.29,0.53}{\textbf{#1}}}

\newcommand{\NormalTok}[1]{#1}

\newcommand{\OtherTok}[1]{\textcolor[rgb]{0.56,0.35,0.01}{#1}}

\newcommand{\SpecialCharTok}[1]{\textcolor[rgb]{0.81,0.36,0.00}{\textbf{#1}}}

\newlength{\cslhangindent}
\newlength{\csllabelwidth}
\newlength{\cslentryspacingunit} 
  {}%
  {\par}
\newenvironment{CSLReferences}[2] 
 {
  \setlength{\parindent}{0pt}
  \ifodd #1
  \let\oldpar\par
  \def\par{\hangindent=\cslhangindent\oldpar}
  \fi
  \setlength{\parskip}{#2\cslentryspacingunit}
 }%
 {}
\usepackage{calc}

\usepackage{booktabs}

\title{LocalCop: An R package for local likelihood inference for
conditional copulas}

        \author[1, 2]{Elif Fidan Acar}
          \author[3]{Martin Lysy}
          \author[1]{Alan Kuchinsky}
    
      \affil[1]{University of Manitoba}
      \affil[2]{Hospital for Sick Children}
      \affil[3]{University of Waterloo}
  \date{\vspace{-5ex}}

\begin{document}
\maketitle

\marginpar{
  \sffamily\small

  {\bfseries DOI:} \href{https://doi.org/}{\color{linky}{}}

  \vspace{2mm}

  {\bfseries Software}
  \begin{itemize}
    \setlength\itemsep{0em}
    \item \href{}{\color{linky}{Review}} \ExternalLink
    \item \href{}{\color{linky}{Repository}} \ExternalLink
    \item \href{}{\color{linky}{Archive}} \ExternalLink
  \end{itemize}

  \vspace{2mm}

  {\bfseries Submitted:} \\
  {\bfseries Published:} 

  \vspace{2mm}
  {\bfseries License}\\
  Authors of papers retain copyright and release the work under a Creative Commons Attribution 4.0 International License (\href{http://creativecommons.org/licenses/by/4.0/}{\color{linky}{CC-BY}}).
}

\newcommand{\R}{{\textsf{R}}}
\newcommand{\cpp}{{\textsf{C++}}}

\hypertarget{summary}{%
\section{Summary}\label{summary}}

Conditional copulas models allow the dependence structure between
multiple response variables to be modelled as a function of covariates.
\textbf{LocalCop} (\protect\hyperlink{ref-localcop}{Acar \& Lysy, 2024})
is an {\textsf{R}}/{\textsf{C++}} package for computationally efficient
semiparametric conditional copula modelling using a local likelihood
inference framework developed in Acar, Craiu, \& Yao
(\protect\hyperlink{ref-ACY2011}{2011}), Acar, Craiu, \& Yao
(\protect\hyperlink{ref-ACY2013}{2013}) and Acar, Czado, \& Lysy
(\protect\hyperlink{ref-ACL2019}{2019}).

\hypertarget{statement-of-need}{%
\section{Statement of Need}\label{statement-of-need}}

There are well-developed {\textsf{R}} packages such as \textbf{copula}
(\protect\hyperlink{ref-copula1}{Hofert, Kojadinovic, Mächler, \& Yan,
2023}; \protect\hyperlink{ref-copula4}{Hofert \& Mächler, 2011};
\protect\hyperlink{ref-copula3}{Kojadinovic \& Yan, 2010};
\protect\hyperlink{ref-copula2}{Yan, 2007}) and \textbf{VineCopula}
(\protect\hyperlink{ref-vinecopula}{Nagler et al., 2023}) for fitting
copulas in various multivariate data settings. However, these software
focus exclusively on unconditional dependence modelling and do not
accommodate covariate information.

Aside from \textbf{LocalCop}, {\textsf{R}} packages for fitting
conditional copulas are \textbf{gamCopula}
(\protect\hyperlink{ref-gamcopula}{Nagler \& Vatter, 2020}) and
\textbf{CondCopulas} (\protect\hyperlink{ref-condcopulas}{Derumigny,
2023}). \textbf{gamCopula} estimates the covariate-dependent copula
parameter using spline smoothing. While this typically has lower
variance than the local likelihood estimate provided by
\textbf{LocalCop}, it also tends to have lower accuracy
(\protect\hyperlink{ref-ACL2019}{Acar et al., 2019}).
\textbf{CondCopulas} estimates the copula parameter using a
semi-parametric maximum-likelihood method based on a kernel-weighted
conditional concordance metric. \textbf{LocalCop} also uses kernel
weighting, but it uses the full likelihood information of a given copula
family rather than just that contained in the concordance metric, and is
therefore more statistically efficient.

Local likelihood methods typically involve solving a large number of
low-dimensional optimization problems and thus can be computationally
intensive. To address this issue, \textbf{LocalCop} implements the local
likelihood function in {\textsf{C++}}, using the
{\textsf{R}}/{\textsf{C++}} package \textbf{TMB}
(\protect\hyperlink{ref-kristensen.etal16}{Kristensen, Nielsen, Berg,
Skaug, \& Bell, 2016}) to efficiently obtain the associated score
function using automatic differentiation. Thus, \textbf{LocalCop} is
able to solve each optimization problem very quickly using
gradient-based algorithms. It also provides a means of easily
parallelizing the optimization across multiple cores, rendering
\textbf{LocalCop} competitive in terms of speed with other available
software for conditional copula estimation.

\hypertarget{background}{%
\section{Background}\label{background}}

For any bivariate response vector \((Y_1, Y_2)\), the conditional joint
distribution given a covariate \(X\) is given by \begin{equation}
F_X(y_1, y_2 \mid x) = C_X (F_{1\mid X} (y_1 \mid x),F_{2\mid X} (y_2 \mid x) \mid x ),
\label{eq:fullmodel}
\end{equation} where \(F_{1\mid X}(y_1 \mid x)\) and
\(F_{2\mid X}(y_2 \mid x)\) are the conditional marginal distributions
of \(Y_1\) and \(Y_2\) given \(X\), and \(C_X(u, v \mid x)\) is a
conditional copula function. That is, for given \(X = x\), the function
\(C_X(u, v \mid x)\) is a bivariate CDF with uniform margins.

The focus of \textbf{LocalCop} is on estimating the conditional copula
function, which is modelled semi-parametrically as \begin{equation}
C_X(u, v \mid x) = \mathcal{C}(u, v\mid \theta(x), \nu),
\label{eq:copmod}
\end{equation} where \(\mathcal{C}(u, v \mid \theta, \nu)\) is one of
the parametric copula families listed in \autoref{tab:calib}, the copula
dependence parameter \(\theta \in \Theta\) is an arbitrary function of
\(X\), and \(\nu \in \Upsilon\) is an additional copula parameter
present in some models. Since most parametric copula families have a
restricted range \(\Theta \subsetneq \mathbb{R}\), we describe the data
generating model (DGM) in terms of the calibration function \(\eta(x)\),
such that \begin{equation}
\theta(x) = g^{-1}(\eta(x)),
\end{equation} where \(g^{-1}: \mathbb{R} \to \Theta\) an inverse-link
function which ensures that the copula parameter has the correct range.
The choice of \(g^{-1}(\eta)\) is not unique and depends on the copula
family. \autoref{tab:calib} displays the copula function
\(\mathcal{C}(u, v \mid \theta, \nu)\) for each of the copula families
provided by \textbf{LocalCop}, along with other relevant information
including the canonical choice of the inverse link function
\(g^{-1}(\eta)\). In \autoref{tab:calib}, \(\Phi^{-1}(p)\) denotes the
inverse CDF of the standard normal; \(t_{\nu}^{-1}(p)\) denotes the
inverse CDF of the Student-t with \(\nu\) degrees of freedom;
\(\Phi_{\theta}(z_1, z_2)\) denotes the CDF of a bivariate normal with
mean \((0, 0)\) and variance
\(\left[\begin{smallmatrix}1 & \theta \\ \theta & 1\end{smallmatrix}\right]\);
and \(t_{\theta,\nu}(z_1, z_2)\) denotes the CDF of a bivariate
Student-t with location \((0, 0)\), scale
\(\left[\begin{smallmatrix}1 & \theta \\ \theta & 1\end{smallmatrix}\right]\),
and degrees of freedom \(\nu\).

\begin{table}

\caption{\label{tab:calib}Copula families implemented in \textbf{LocalCop}.}
\centering
\resizebox{\linewidth}{!}{
\begin{tabular}[t]{llllll}
\toprule
Family & $\mathcal{C}(u, v \mid \theta,\nu)$ & $\theta \in \Theta$ & $\nu \in \Upsilon$ & $g^{-1}(\eta)$ & $\tau(\theta)$\\
\midrule
Gaussian & $\Phi_\theta ( \Phi^{-1}(u), \Phi^{-1}(v))$ & $(-1,1)$ & - & $\dfrac{e^{\eta} - e^{-\eta}}{e^{\eta} + e^{-\eta}}$ & $\frac{2}{\pi}\arcsin(\theta)$\\
Student-t & $t_{\theta,\nu} ( t_\nu^{-1}(u), t_\nu^{-1}(v))$ & $(-1,1)$ & $(0, \infty)$ & $\dfrac{e^{\eta} - e^{-\eta}}{e^{\eta} + e^{-\eta}}$ & $\frac{2}{\pi}\arcsin(\theta)$\\
Clayton & $\displaystyle (u^{-\theta} + v^{-\theta} -1)^{-\frac{1}{\theta}}$ & $(0, \infty)$ & - & $e^\eta$ & $\frac{\theta}{\theta + 2}$\\
Gumbel & $\displaystyle  \exp\left[ - \{  (-\log u)^\theta + (-\log v)^\theta\}^{\frac{1}{\theta}} \right]$ & $[1, \infty)$ & - & $e^\eta + 1$ & $1 - \frac{1}{\theta}$\\
Frank & $-\frac{1}{\theta}\log \left\{ 1 + \frac{(e^{-\theta u} - 1)(e^{-\theta v} - 1)}{e^{-\theta} - 1}\right\}$ & $(-\infty, \infty)\setminus\{0\}$ & - & $\eta$ & no closed form\\
\bottomrule
\end{tabular}}
\end{table}

Local likelihood estimation of the conditional copula parameter
\(\theta(x)\) uses Taylor expansions to approximate the calibration
function \(\eta(x)\) at an observed covariate value \(X = x\) near a
fixed point \(X = x_0\), i.e., \[
\eta(x)\approx \eta(x_0) + \eta^{(1)}(x_0) (x - x_0) + \ldots + \dfrac{\eta^{(p)}(x_0)}{p!} (x - x_0)^{p}.
\] One then estimates \(\beta_k = \eta^{(k)}(x_0)/k!\) for
\(k = 0,\ldots,p\) using a kernel-weighted local likelihood function
\begin{equation}
\ell(\boldsymbol{\beta}) = \sum_{i=1}^n \log\left\{ c\left(u_i, v_i \mid g^{-1}( \boldsymbol{x}_{i}^T \boldsymbol{\beta}), \nu \right)\right\} K_h\left(\dfrac{x_i-x_0}{h}\right),
\label{eq:locallik}
\end{equation} where \((u_i, v_i, x_i)\) is the data for observation
\(i\),
\(\boldsymbol{x}_i = (1, x_i - x_0, (x_i - x_0)^2, \ldots, (x_i - x_0)^p)\),
\(\boldsymbol{\beta}= (\beta_0, \beta_1, \ldots, \beta_p)\), and
\(K_h(z)\) is a kernel function with bandwidth parameter \(h > 0\).
Having maximized \(\ell(\boldsymbol{\beta})\) in \autoref{eq:locallik},
one estimates \(\eta(x_0)\) by \(\hat \eta(x_0) = \hat \beta_0\).
Usually, a linear fit with \(p=1\) suffices to obtain a good estimate in
practice.

\hypertarget{usage}{%
\section{Usage}\label{usage}}

\textbf{LocalCop} is available on
\href{https://CRAN.R-project.org/package=LocalCop}{CRAN} and
\href{https://github.com/mlysy/LocalCop}{GitHub}. The two main package
functions are:

\begin{itemize}
\item
  \texttt{CondiCopLocFit()}: For estimating the calibration function at
  a sequence of values \(\boldsymbol{x}_0 = (x_{01}, \ldots, x_{0m})\).
\item
  \texttt{CondiCopSelect()}: For selecting a copula family and bandwidth
  parameter using leave-one-out cross-validation (LOO-CV) with
  subsampling as described in Acar et al.
  (\protect\hyperlink{ref-ACL2019}{2019}).
\end{itemize}

\begin{figure}
\includegraphics[width=1\linewidth]{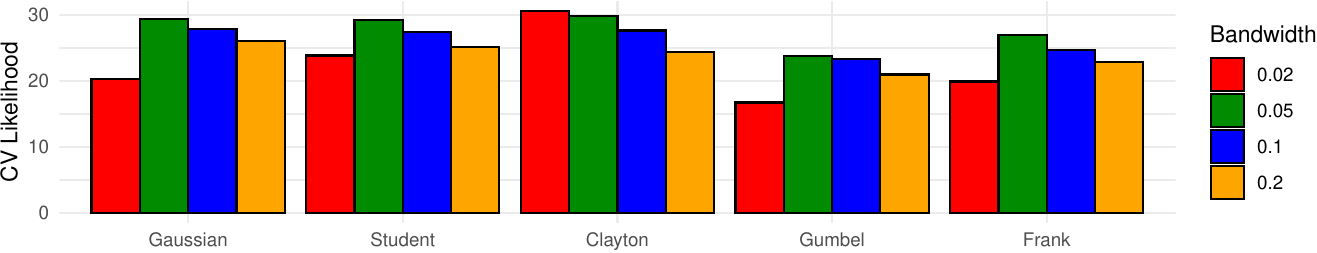} \caption{Cross-validation metric for each combination of family and bandwidth.}\label{fig:select1-plot}
\end{figure}

\begin{figure}
\includegraphics[width=1\linewidth]{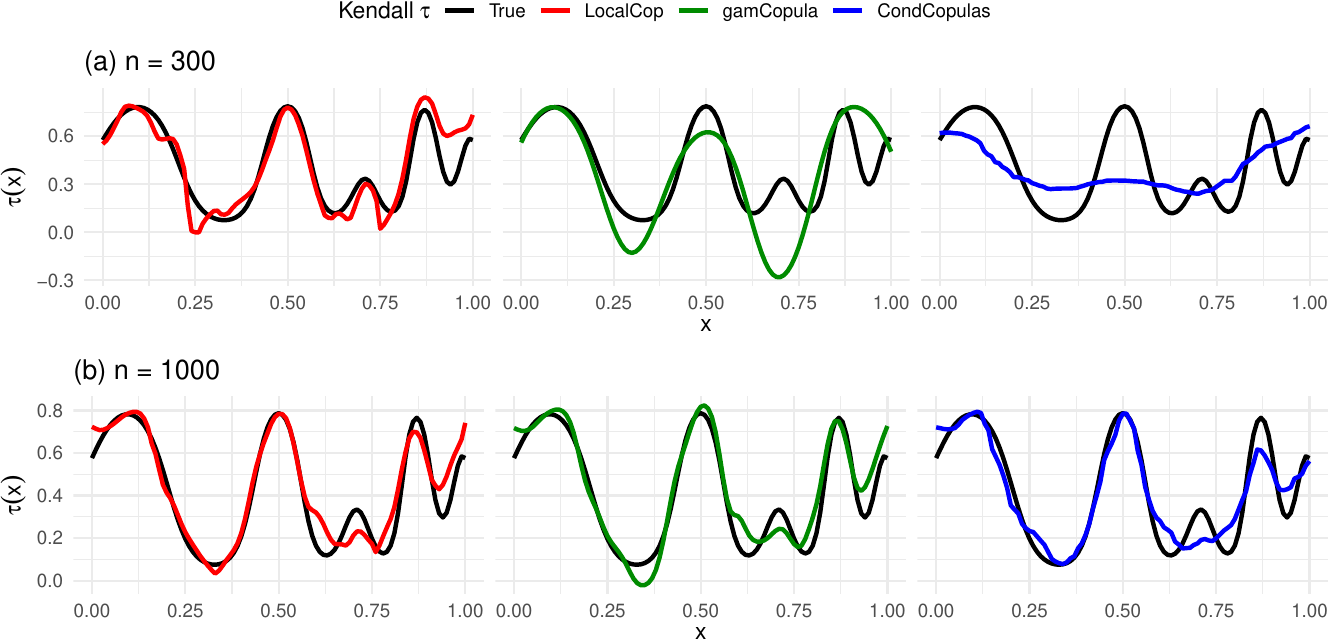} \caption{True vs estimated conditional Kendall $\tau$ using various methods.}\label{fig:copcomp}
\end{figure}

In the following example, we illustrate the model selection/tuning and
fitting steps for data generated from a Clayton copula with conditional
Kendall \(\tau\) displayed in \autoref{fig:copcomp}. The CV metric for
each combination of family and bandwidth are displayed in
\autoref{fig:select1-plot}.

\begin{Shaded}
\begin{Highlighting}[]
\FunctionTok{library}\NormalTok{(LocalCop)   }\CommentTok{\# local likelihood estimation}
\FunctionTok{library}\NormalTok{(VineCopula) }\CommentTok{\# simulate copula data}
\end{Highlighting}
\end{Shaded}

\begin{Shaded}
\begin{Highlighting}[]
\FunctionTok{set.seed}\NormalTok{(}\DecValTok{2024}\NormalTok{)}

\CommentTok{\# simulation setting}
\NormalTok{family }\OtherTok{\textless{}{-}} \DecValTok{3}                    \CommentTok{\# Clayton Copula}
\NormalTok{n\_obs }\OtherTok{\textless{}{-}} \DecValTok{300}                   \CommentTok{\# number of observations}
\NormalTok{eta\_fun }\OtherTok{\textless{}{-}} \ControlFlowTok{function}\NormalTok{(x) \{       }\CommentTok{\# calibration function}
  \FunctionTok{sin}\NormalTok{(}\DecValTok{5}\SpecialCharTok{*}\NormalTok{pi}\SpecialCharTok{*}\NormalTok{x) }\SpecialCharTok{+} \FunctionTok{cos}\NormalTok{(}\DecValTok{8}\SpecialCharTok{*}\NormalTok{pi}\SpecialCharTok{*}\NormalTok{x}\SpecialCharTok{\^{}}\DecValTok{2}\NormalTok{)}
\NormalTok{\}}
\end{Highlighting}
\end{Shaded}

\begin{Shaded}
\begin{Highlighting}[]
\CommentTok{\# simulate covariate values}
\NormalTok{x }\OtherTok{\textless{}{-}} \FunctionTok{sort}\NormalTok{(}\FunctionTok{runif}\NormalTok{(n\_obs))}

\CommentTok{\# simulate response data}
\NormalTok{eta\_true }\OtherTok{\textless{}{-}} \FunctionTok{eta\_fun}\NormalTok{(x)                     }\CommentTok{\# calibration parameter eta(x)}
\NormalTok{par\_true }\OtherTok{\textless{}{-}} \FunctionTok{BiCopEta2Par}\NormalTok{(}\AttributeTok{family =}\NormalTok{ family,  }\CommentTok{\# copula parameter theta(x)}
                         \AttributeTok{eta =}\NormalTok{ eta\_true)}
\NormalTok{udata }\OtherTok{\textless{}{-}}\NormalTok{ VineCopula}\SpecialCharTok{::}\FunctionTok{BiCopSim}\NormalTok{(n\_obs, }\AttributeTok{family =}\NormalTok{ family, }\AttributeTok{par =}\NormalTok{ par\_true)}
\end{Highlighting}
\end{Shaded}

\begin{Shaded}
\begin{Highlighting}[]
\CommentTok{\# model selection and tuning}
\NormalTok{bandset }\OtherTok{\textless{}{-}} \FunctionTok{c}\NormalTok{(.}\DecValTok{02}\NormalTok{, .}\DecValTok{05}\NormalTok{, .}\DecValTok{1}\NormalTok{, .}\DecValTok{2}\NormalTok{) }\CommentTok{\# set of bandwidth parameters}
\NormalTok{famset }\OtherTok{\textless{}{-}} \FunctionTok{c}\NormalTok{(}\DecValTok{1}\NormalTok{, }\DecValTok{2}\NormalTok{, }\DecValTok{3}\NormalTok{, }\DecValTok{4}\NormalTok{, }\DecValTok{5}\NormalTok{)     }\CommentTok{\# set of copula families}
\NormalTok{kernel }\OtherTok{\textless{}{-}}\NormalTok{ KernGaus             }\CommentTok{\# kernel function}
\NormalTok{degree }\OtherTok{\textless{}{-}} \DecValTok{1}                    \CommentTok{\# degree of local polynomial}
\NormalTok{n\_loo }\OtherTok{\textless{}{-}} \DecValTok{100}                   \CommentTok{\# number of LOO{-}CV observations}
                               \CommentTok{\# (can be much smaller than n\_obs)}
\end{Highlighting}
\end{Shaded}

\begin{Shaded}
\begin{Highlighting}[]
\CommentTok{\# calculate cv for each combination of family and bandwidth}
\NormalTok{cvselect }\OtherTok{\textless{}{-}} \FunctionTok{CondiCopSelect}\NormalTok{(}\AttributeTok{u1=}\NormalTok{ udata[,}\DecValTok{1}\NormalTok{], }\AttributeTok{u2 =}\NormalTok{ udata[,}\DecValTok{2}\NormalTok{],}
                           \AttributeTok{x =}\NormalTok{ x, }\AttributeTok{xind =}\NormalTok{ n\_loo,}
                           \AttributeTok{kernel =}\NormalTok{ kernel, }\AttributeTok{degree =}\NormalTok{ degree,}
                           \AttributeTok{family =}\NormalTok{ famset, }\AttributeTok{band =}\NormalTok{ bandset)}
\end{Highlighting}
\end{Shaded}

\begin{Shaded}
\begin{Highlighting}[]
\CommentTok{\# extract the selected family and bandwidth from cvselect}
\NormalTok{cv\_res }\OtherTok{\textless{}{-}}\NormalTok{ cvselect}\SpecialCharTok{$}\NormalTok{cv}
\NormalTok{i\_opt }\OtherTok{\textless{}{-}} \FunctionTok{which.max}\NormalTok{(cv\_res}\SpecialCharTok{$}\NormalTok{cv)}
\NormalTok{fam\_opt }\OtherTok{\textless{}{-}}\NormalTok{ cv\_res[i\_opt,]}\SpecialCharTok{$}\NormalTok{family}
\NormalTok{band\_opt }\OtherTok{\textless{}{-}}\NormalTok{ cv\_res[i\_opt,]}\SpecialCharTok{$}\NormalTok{band}

\CommentTok{\# calculate eta(x) on a grid of values}
\NormalTok{x0 }\OtherTok{\textless{}{-}} \FunctionTok{seq}\NormalTok{(}\DecValTok{0}\NormalTok{, }\DecValTok{1}\NormalTok{, }\AttributeTok{by =} \FloatTok{0.01}\NormalTok{)}
\NormalTok{copfit }\OtherTok{\textless{}{-}} \FunctionTok{CondiCopLocFit}\NormalTok{(}\AttributeTok{u1 =}\NormalTok{ udata[,}\DecValTok{1}\NormalTok{], }\AttributeTok{u2 =}\NormalTok{ udata[,}\DecValTok{2}\NormalTok{],}
                         \AttributeTok{x =}\NormalTok{ x, }\AttributeTok{x0 =}\NormalTok{ x0,}
                         \AttributeTok{kernel =}\NormalTok{ kernel, }\AttributeTok{degree =}\NormalTok{ degree,}
                         \AttributeTok{family =}\NormalTok{ fam\_opt, }\AttributeTok{band =}\NormalTok{ band\_opt)}
\CommentTok{\# convert eta to Kendall tau}
\NormalTok{tau\_loc }\OtherTok{\textless{}{-}} \FunctionTok{BiCopEta2Tau}\NormalTok{(copfit}\SpecialCharTok{$}\NormalTok{eta, }\AttributeTok{family=}\NormalTok{ fam\_opt)}
\end{Highlighting}
\end{Shaded}

\begin{Shaded}
\begin{Highlighting}[]
\CommentTok{\# simulate covariate values}
\NormalTok{x }\OtherTok{\textless{}{-}} \FunctionTok{sort}\NormalTok{(}\FunctionTok{runif}\NormalTok{(n\_obs))}

\CommentTok{\# simulate response data}
\NormalTok{eta\_true }\OtherTok{\textless{}{-}} \FunctionTok{eta\_fun}\NormalTok{(x)                     }\CommentTok{\# calibration parameter eta(x)}
\NormalTok{par\_true }\OtherTok{\textless{}{-}} \FunctionTok{BiCopEta2Par}\NormalTok{(}\AttributeTok{family =}\NormalTok{ family,  }\CommentTok{\# copula parameter theta(x)}
                         \AttributeTok{eta =}\NormalTok{ eta\_true)}
\NormalTok{udata }\OtherTok{\textless{}{-}}\NormalTok{ VineCopula}\SpecialCharTok{::}\FunctionTok{BiCopSim}\NormalTok{(n\_obs, }\AttributeTok{family =}\NormalTok{ family, }\AttributeTok{par =}\NormalTok{ par\_true)}
\end{Highlighting}
\end{Shaded}

\begin{Shaded}
\begin{Highlighting}[]
\CommentTok{\# model selection and tuning}
\NormalTok{bandset }\OtherTok{\textless{}{-}} \FunctionTok{c}\NormalTok{(.}\DecValTok{02}\NormalTok{, .}\DecValTok{05}\NormalTok{, .}\DecValTok{1}\NormalTok{, .}\DecValTok{2}\NormalTok{) }\CommentTok{\# set of bandwidth parameters}
\NormalTok{famset }\OtherTok{\textless{}{-}} \FunctionTok{c}\NormalTok{(}\DecValTok{1}\NormalTok{, }\DecValTok{2}\NormalTok{, }\DecValTok{3}\NormalTok{, }\DecValTok{4}\NormalTok{, }\DecValTok{5}\NormalTok{)     }\CommentTok{\# set of copula families}
\NormalTok{kernel }\OtherTok{\textless{}{-}}\NormalTok{ KernGaus             }\CommentTok{\# kernel function}
\NormalTok{degree }\OtherTok{\textless{}{-}} \DecValTok{1}                    \CommentTok{\# degree of local polynomial}
\NormalTok{n\_loo }\OtherTok{\textless{}{-}} \DecValTok{100}                   \CommentTok{\# number of LOO{-}CV observations}
                               \CommentTok{\# (can be much smaller than n\_obs)}
\end{Highlighting}
\end{Shaded}

\begin{Shaded}
\begin{Highlighting}[]
\CommentTok{\# calculate cv for each combination of family and bandwidth}
\NormalTok{cvselect }\OtherTok{\textless{}{-}} \FunctionTok{CondiCopSelect}\NormalTok{(}\AttributeTok{u1=}\NormalTok{ udata[,}\DecValTok{1}\NormalTok{], }\AttributeTok{u2 =}\NormalTok{ udata[,}\DecValTok{2}\NormalTok{],}
                           \AttributeTok{x =}\NormalTok{ x, }\AttributeTok{xind =}\NormalTok{ n\_loo,}
                           \AttributeTok{kernel =}\NormalTok{ kernel, }\AttributeTok{degree =}\NormalTok{ degree,}
                           \AttributeTok{family =}\NormalTok{ famset, }\AttributeTok{band =}\NormalTok{ bandset)}
\end{Highlighting}
\end{Shaded}

\begin{Shaded}
\begin{Highlighting}[]
\CommentTok{\# extract the selected family and bandwidth from cvselect}
\NormalTok{cv\_res }\OtherTok{\textless{}{-}}\NormalTok{ cvselect}\SpecialCharTok{$}\NormalTok{cv}
\NormalTok{i\_opt }\OtherTok{\textless{}{-}} \FunctionTok{which.max}\NormalTok{(cv\_res}\SpecialCharTok{$}\NormalTok{cv)}
\NormalTok{fam\_opt }\OtherTok{\textless{}{-}}\NormalTok{ cv\_res[i\_opt,]}\SpecialCharTok{$}\NormalTok{family}
\NormalTok{band\_opt }\OtherTok{\textless{}{-}}\NormalTok{ cv\_res[i\_opt,]}\SpecialCharTok{$}\NormalTok{band}

\CommentTok{\# calculate eta(x) on a grid of values}
\NormalTok{x0 }\OtherTok{\textless{}{-}} \FunctionTok{seq}\NormalTok{(}\DecValTok{0}\NormalTok{, }\DecValTok{1}\NormalTok{, }\AttributeTok{by =} \FloatTok{0.01}\NormalTok{)}
\NormalTok{copfit }\OtherTok{\textless{}{-}} \FunctionTok{CondiCopLocFit}\NormalTok{(}\AttributeTok{u1 =}\NormalTok{ udata[,}\DecValTok{1}\NormalTok{], }\AttributeTok{u2 =}\NormalTok{ udata[,}\DecValTok{2}\NormalTok{],}
                         \AttributeTok{x =}\NormalTok{ x, }\AttributeTok{x0 =}\NormalTok{ x0,}
                         \AttributeTok{kernel =}\NormalTok{ kernel, }\AttributeTok{degree =}\NormalTok{ degree,}
                         \AttributeTok{family =}\NormalTok{ fam\_opt, }\AttributeTok{band =}\NormalTok{ band\_opt)}
\CommentTok{\# convert eta to Kendall tau}
\NormalTok{tau\_loc }\OtherTok{\textless{}{-}} \FunctionTok{BiCopEta2Tau}\NormalTok{(copfit}\SpecialCharTok{$}\NormalTok{eta, }\AttributeTok{family=}\NormalTok{ fam\_opt)}
\end{Highlighting}
\end{Shaded}

In \autoref{fig:copcomp}, we compare the true conditional Kendall
\(\tau\) to estimates using each of the three conditional copula fitting
packages \textbf{LocalCop}, \textbf{gamCopula}, and
\textbf{CondCopulas}, for sample sizes \(n = 300\) and \(n = 1000\). In
\textbf{gamCopula}, selection of the copula family smoothing splines is
done using the generalized CV framework provided by the
{\textsf{R}} package \textbf{mgcv} (\protect\hyperlink{ref-wood17}{Wood,
2017}). In \textbf{CondCopulas}, selection of the bandwidth parameter is
done using LOO-CV. In this particular example, the sample size of
\(n = 300\) is not large enough for \textbf{gamCopula} to pick a
sufficiently flexible spline basis, and \textbf{CondCopulas} picks a
large bandwidth which oversmooths the data. For the larger sample size
\(n = 1000\), the three methods exhibit similar accuracy.

\hypertarget{acknowledgements}{%
\section{Acknowledgements}\label{acknowledgements}}

We acknowledge funding support from the Natural Sciences and Engineering
Research Council of Canada Discovery Grants RGPIN-2020-06753 (Acar) and
RGPIN-2020-04364 (Lysy).

\hypertarget{references}{%
\section*{References}\label{references}}
\addcontentsline{toc}{section}{References}

\hypertarget{refs}{}
\begin{CSLReferences}{1}{0}
\leavevmode\vadjust pre{\hypertarget{ref-ACY2011}{}}%
Acar, E. F., Craiu, R. V., \& Yao, F. (2011). Dependence calibration in
conditional copulas: A nonparametric approach. \emph{Biometrics},
\emph{67}(2), 445--453.

\leavevmode\vadjust pre{\hypertarget{ref-ACY2013}{}}%
Acar, E. F., Craiu, R. V., \& Yao, F. (2013). Statistical testing of
covariate effects in conditional copula models. \emph{Electronic Journal
of Statistics}, \emph{7}, 2822--2850.

\leavevmode\vadjust pre{\hypertarget{ref-ACL2019}{}}%
Acar, E. F., Czado, C., \& Lysy, M. (2019). Dynamic vine copula models
for multivariate time series data. \emph{Econometrics and Statistics},
\emph{12}, 181--197.

\leavevmode\vadjust pre{\hypertarget{ref-localcop}{}}%
Acar, E. F., \& Lysy, M. (2024). \emph{LocalCop: LocalCop: Local
likelihood inference for conditional copula models}. Retrieved from
\url{https://CRAN.R-project.org/package=LocalCop}

\leavevmode\vadjust pre{\hypertarget{ref-condcopulas}{}}%
Derumigny, A. (2023). \emph{CondCopulas: Estimation and inference for
conditional copula models}. Retrieved from
\url{https://CRAN.R-project.org/package=CondCopulas}

\leavevmode\vadjust pre{\hypertarget{ref-copula1}{}}%
Hofert, M., Kojadinovic, I., Mächler, M., \& Yan, J. (2023).
\emph{Copula: Multivariate dependence with copulas}. Retrieved from
\url{https://CRAN.R-project.org/package=copula}

\leavevmode\vadjust pre{\hypertarget{ref-copula4}{}}%
Hofert, M., \& Mächler, M. (2011). Nested archimedean copulas meet {R}:
The {nacopula} package. \emph{Journal of Statistical Software},
\emph{39}(9), 1--20. Retrieved from
\url{https://www.jstatsoft.org/v39/i09/}

\leavevmode\vadjust pre{\hypertarget{ref-copula3}{}}%
Kojadinovic, I., \& Yan, J. (2010). Modeling multivariate distributions
with continuous margins using the {copula} {R} package. \emph{Journal of
Statistical Software}, \emph{34}(9), 1--20. Retrieved from
\url{https://www.jstatsoft.org/v34/i09/}

\leavevmode\vadjust pre{\hypertarget{ref-kristensen.etal16}{}}%
Kristensen, K., Nielsen, A., Berg, C. W., Skaug, H., \& Bell, B. M.
(2016). {TMB}: Automatic differentiation and {L}aplace approximation.
\emph{Journal of Statistical Software}, \emph{70}(5), 1--21.
doi:\href{https://doi.org/10.18637/jss.v070.i05}{10.18637/jss.v070.i05}

\leavevmode\vadjust pre{\hypertarget{ref-vinecopula}{}}%
Nagler, T., Schepsmeier, U., Stoeber, J., Brechmann, E. C., Graeler, B.,
\& Erhardt, T. (2023). \emph{VineCopula: Statistical inference of vine
copulas}. Retrieved from
\url{https://CRAN.R-project.org/package=VineCopula}

\leavevmode\vadjust pre{\hypertarget{ref-gamcopula}{}}%
Nagler, T., \& Vatter, T. (2020). \emph{gamCopula: Generalized additive
models for bivariate conditional dependence structures and vine
copulas}. Retrieved from
\url{https://CRAN.R-project.org/package=gamCopula}

\leavevmode\vadjust pre{\hypertarget{ref-wood17}{}}%
Wood, S. N. (2017). \emph{Generalized additive models: An introduction
with {R}} (2nd ed.). Chapman; Hall/CRC.

\leavevmode\vadjust pre{\hypertarget{ref-copula2}{}}%
Yan, J. (2007). Enjoy the joy of copulas: With a package {copula}.
\emph{Journal of Statistical Software}, \emph{21}(4), 1--21. Retrieved
from \url{https://www.jstatsoft.org/v21/i04/}

\end{CSLReferences}

\end{document}